\documentclass[aps,prd,reprint,longbibliography,superscriptaddress,nofootinbib,floatfix]{revtex4-1}
\pdfoutput=1

\usepackage{amsfonts,amssymb,amsmath}
\usepackage{epsfig}
\usepackage{bm}
\usepackage{latexsym}
\usepackage{natbib}
\usepackage{url}
\usepackage{dcolumn}
\usepackage{color}
\usepackage{graphicx,epsfig}
\usepackage{hyperref}
\hypersetup{colorlinks   = true,
            urlcolor     = blue,
            citecolor    = blue,
            linkcolor    = blue,
            menucolor    = blue,
            anchorcolor  = blue,
            filecolor    = blue
}

\enlargethispage{\baselineskip}

\smallskipamount

\everymath{\displaystyle}

\newcommand{\TDLI}{\affiliation{Tsung-Dao Lee Institute (TDLI), No.\ 1 Lisuo Road, 201210 Shanghai, China}}
\newcommand{\SJTU}{\affiliation{School of Physics and Astronomy, Shanghai Jiao Tong University, \\ Dongchuan Road 800, 200240 Shanghai, China}}
\newcommand{\UNISA}{\affiliation{Dipartimento di Fisica ``E.R.\ Caianiello'', Universit\`a degli Studi di Salerno,\\ Via Giovanni Paolo II, 132 - 84084 Fisciano (SA), Italy}}
\newcommand{\INFNSA}{\affiliation{Istituto Nazionale di Fisica Nucleare - Gruppo Collegato di Salerno - Sezione di Napoli,\\ Via Giovanni Paolo II, 132 - 84084 Fisciano (SA), Italy}}
\newcommand{\INFNPI}{\affiliation{Istituto Nazionale di Fisica Nucleare, Sezione di Pisa, Largo Bruno Pontecorvo 3, I-56127 Pisa, Italy}}

\begin{document}

\title{Ultralight Black Holes as Sources of High-Energy Particles}

\author{Michael Zantedeschi}
\email{michael.zantedeschi@pi.infn.it}
\TDLI \SJTU \INFNPI
\author{Luca Visinelli}
\email{lvisinelli@unisa.it \newline }
\UNISA \INFNSA

\begin{abstract}
The \textit{memory burden} effect, the idea that the amount of information stored within a system contributes to its stabilization, is particularly relevant for systems with a large information storage capacity, such as black holes. In these objects, the evaporation process halts, at the latest, once approximately half of the initial mass has been radiated away. As a result, light primordial black holes (PBHs) with mass $m_{\rm PBH} \lesssim 10^{15}\,\mathrm{g}$, which are traditionally assumed to have fully evaporated by the present time, may instead survive and constitute viable dark matter candidates.
Ongoing mergers of such PBHs would give rise to ``young'' black holes that resume their evaporation, emitting ultrahigh-energy particles potentially detectable by current experiments. The resulting emission spectrum would be thermal across all Standard Model particle species, offering a clear and distinctive signature. 
We demonstrate that, if the memory burden effect activates after PBHs have lost around half of their initial mass, current measurements of the neutrino flux at Earth place strong constraints on such dark matter candidates for $m_{\rm PBH} \lesssim 10^9\,\mathrm{g}$. This suggests that the memory burden must set in at earlier stages of evaporation.
Unlike existing bounds, our results depend solely on the mass of the remnant, and not on model-dependent details of the stabilized phase. We also discuss the potential for refining these constraints through observations of gamma rays, cosmic rays, and gravitational waves.
\end{abstract}

\maketitle

\section{Introduction}

Black holes (BHs), predicted by general relativity, are not only fascinating astrophysical objects but also intriguing sources of particle emission. A BH of mass $m_{\rm BH}$ emits high-energy particles lighter than its Hawking temperature, $T_{\rm H} = (8\pi G m_{\rm BH})^{-1}$~\cite{Hawking:1975vcx}. In the semiclassical regime where quantum gravity effects are neglected throughout their evaporation, the mass loss rate for an emitting BH follows $\dot m_{\rm BH} \propto m_{\rm BH}^{-2}$. The Hawking radiation has been searched for in terms of the BH emission into photons~\cite{Carr:2016hva, Boudaud:2018hqb, DeRocco:2019fjq, Laha:2019ssq, Laha:2020ivk, Chan:2020zry}, neutrinos~\cite{Dasgupta:2019cae,Bernal:2022swt}, gravitons~\cite{Hooper:2020evu}, and exotic particles~\cite{Hooper:2019gtx, Schiavone:2021imu, Auffinger:2022khh, Baker:2022rkn}.

However, it has been suggested that the information contained within a BH affects its decay, slowing down the evaporation process through a ``memory burden'' effect~\cite{Dvali:2018xpy,Dvali:2020wft,Dvali:2024hsb}, which becomes significant after a portion of its mass has evaporated~\cite{Dvali:2020wft,Dvali:2024hsb}. The timescale that signals the breakdown of the semiclassical description and the set in of memory burden is
\begin{equation}
    \label{eq:tausc}
    \tau_{\rm sc} \simeq q\,\tau_{\rm evap}\,,
\end{equation}
where $\tau_{\rm evap}$ is the BH evaporation time according to the semiclassical rate and the prefactor accounts for the fraction $q\doteq \Delta m_{\rm PBH}/m_{\rm PBH}$ of the BH mass being lost during the semiclassical emission phase. Here, we fix $q=1/2$, corresponding to the latest time that memory burden may kick in.
The subsequent dynamics of the memory-burdened BH for times $t \gtrsim \tau_{\rm sc}$ remain uncertain due to the influence of quantum gravity effects~\cite{Dvali:2020wft,Dvali:2024hsb}. One possible outcome is that the configuration becomes unstable and eventually breaks down. Alternatively, evaporation could persist at a reduced rate, controlled by powers of the gravitational coupling, $\alpha_{\rm gr} \simeq S^{-1}$. In what follows, we assume the former scenario as the likely outcome which is further indicated by theoretical analyses~\cite{Dvali:2020wft,Dvali:2024hsb}. 

When quantum effects are taken into account, the BH evaporation rate is slowed down and the BH lifetime is extended as~\cite{Dvali:2020wft}
\begin{equation}
\label{eq:lifetimeenhancement}
    \tau \simeq \tau_{\rm evap} \,S^{k}\,,
\end{equation}
where $S = 4\pi G m_{\rm BH}^2$ represents the BH entropy. Theoretical models such as the holographic description of BHs~\cite{Dvali:2017nis} and the microscopic description in terms of marginally bound gravitons—the so-called BH $N$-portrait~\cite{Dvali:2011aa, Dvali:2012en, Dvali:2013eja}- independently suggest that $k = 2$~\cite{Dvali:2020wft,Dvali:2024hsb} (consistency requires $k\geq 1$). In what follows, we will derive bounds independent of this quantity.

The effects described above are particularly relevant for BHs that have formed primordially, or PBHs. A PBH emitting semiclassically has not fully evaporated to date if its mass is $m_{\rm PBH} \gtrsim 10^{15}$\,g. As a result, PBHs have long been considered potential dark matter (DM) candidates~\cite{Zeldovich:1967lct, Hawking:1971ei, Carr:1974nx} (see e.g., \cite{Escriva:2022duf} for a review). Even if PBHs constitute only a fraction $f_{\rm PBH}$ of the present DM abundance, they could still reveal their existence through a variety of astrophysical~\cite{Pani:2014rca, Boucenna:2017ghj, Smyth:2019whb, Carr:2020erq, Carr:2020mqm, Saha:2021pqf,Tran:2023jci,Saha:2024ies,Thoss:2024dkg}, particle~\cite{Lunardini:2019zob}, and cosmological phenomena~\cite{Kohri:1999ex, Baumann:2007yr, Carr:2009jm, Hook:2014mla, Fujita:2014hha, Hooper:2020otu, Perez-Gonzalez:2020vnz, Smyth:2021lkn, Mazde:2022sdx, Bernal:2022pue, Calabrese:2023key, Calabrese:2023bxz, Gehrman:2022imk, Borah:2024qyo, Boccia:2024nly}. 

The memory burden effect introduces a new mass window for PBHs as DM candidates below $10^{15}$\,g, with BHs of mass around $10^{8}$\,g shown to be viable while satisfying all known constraints~\cite{Dvali:2020wft}. For $k=2$, Eq.~\eqref{eq:lifetimeenhancement} implies that memory-burdened BHs as light as $\gtrsim  10^3\rm g$ can be DM candidates. A viable mechanism for their production within this mass range $3\times 10^4 \lesssim m_{\rm PBH}/{\rm g} \lesssim 10^9$, here dubbed the \textit{ultralight} BH window, was introduced in Ref.~\cite{Dvali:2021byy}.\footnote{See also Refs.~\cite{Carr:1994ar, Profumo:2024fxq} for Planck-scale relics.} A more precise mass range for memory-burdened PBHs constituting DM has been recently derived~\cite{Alexandre:2024nuo,Thoss:2024hsr}. Other consequences of memory-burdened BHs have also been recently explored~\cite{Balaji:2024hpu, Haque:2024eyh, Riotto:2024ayo, Barman:2024iht, Barman:2024ufm, Bhaumik:2024qzd, Kohri:2024qpd, Jiang:2024aju, Chianese:2024rsn, Athron:2024fcj, Dvali:2025ktz}.

In this paper, we discuss the observational consequences of memory-burdened PBHs as DM. We argue that these PBHs, locked in binary configurations, merge to form new BHs of similar mass. Upon merging, the new object is a young BH, no longer subject to the memory burden effect. As such, the freshly-formed BH starts emitting again for a duration $\tau_{\rm sc}$, at much lower redshifts than their progenitors, thus leading to novel signatures that are unique to the memory burden phenomenon. 

Memory-burdened PBH mergers could potentially occur within our own Galaxy if these objects constitute a significant DM fraction. The light BHs resulting from their mergers would be bright sources of gamma and cosmic rays, detectable at observatories~\cite{Coleman:2022abf}, as well as sources of high and ultra high-energy neutrinos detectable by neutrino telescopes~\cite{IceCube:2010tuw, IceCube:2011fxx}. Gravitons in ultra-high frequency windows are also emitted, providing new target opportunities for detection facilities. Here, we work with natural units $\hbar=c=1$.

\section{Methods}

\subsection{Memory burden effect} Saturated configurations offer valuable insights into objects with high information-storage capacity, revealing connections between gravitational and non-gravitational systems. Notably, ``saturons’’, configurations of maximal entropy constrained by the unitarity bound~\cite{Dvali:2020wqi}, exemplify these objects. BHs are a prime example of saturated configurations, as they obey the entropy-area law~\cite{Bekenstein:1973ur}. Saturons also arise in non-gravitational theories, such as renormalizable field theories~\cite{Dvali:2020wqi, Dvali:2019jjw, Dvali:2019ulr}. These objects not only follow an area law, but also share several properties with BHs, such as thermal-like evaporation~\cite{Dvali:2021rlf, Dvali:2021tez}, the presence of an information horizon in the semiclassical regime~\cite{Dvali:2019jjw, Dvali:2019ulr, Dvali:2020wqi, Dvali:2021jto, Dvali:2021rlf, Dvali:2021tez}, a timescale for information retrieval corresponding to the half-time of evaporation~\cite{Dvali:2020wqi, Dvali:2021rlf, Dvali:2021tez}, and a maximal spin bound that cannot exceed their entropy, similar to the extremality condition for spinning BHs~\cite{Dvali:2021ofp, Dvali:2023qlk}. As a consequence, many features traditionally associated with BHs may arise from the broader behavior of saturated configurations, rather than being unique to gravity.

This framework provides new approaches for studying BHs from both theoretical and phenomenological perspectives. By analyzing the non-gravitational counterparts of BHs, we can use the calculability of renormalizable theories to better understand their enigmatic properties. Moreover, the study of this class of objects, allows for the extension of novel properties to BHs by universality. 

The memory burden phenomenon is a necessary feature of all saturons, including BHs. This stems from the requirement of generating a large microstate degeneracy to account for the entropy-area law. 
Achieving this requires a large number of modes with zero energy gap or ``gapless'', 
localized on the support of the BH.
Outside the object, these same degrees of freedom become gapped. As a result, for the BH to release its memory, an energy barrier needs to be overcome. Hawking radiation cannot do this due to its softness. The very same energy barrier is the one the BH eventually has to face as Hawking evaporation reduces its mass. We shall not rederive the memory burden mechanism here, and instead refer the reader to the relevant literature~\cite{Dvali:2020wqi, Dvali:2019jjw, Dvali:2019ulr}. Our focus is on features relevant for the merger process.

The memory stored in a BH can be represented as $M$ qubits, which, due to their gaplessness, leads to a number of degenerate microstates $n_{\rm st}= 2^M$, and therefore an entropy $S=\log n_{\rm st}\simeq M$. The amount of memory stored within the BH, corresponding to the total number of occupied memory modes $N_{\rm G}$, is determined statistically, according to the probability distribution~\cite{Dvali:2024hsb},
\begin{equation}
    \label{eq:probability}
    {\mathcal P}_{N_{\rm G}} \,  = \, 2^{-S} \frac{S!}{(S-N_{\rm G})! N_{\rm G}!} \,,
\end{equation}
where $N_{\rm G}$ is effectively the order parameter responsible for the quantum backreaction~\cite{Dvali:2024hsb}. On average, $N_{\rm G}\simeq S/2$. As BH decays, the increasing mass gap of the memory modes eventually halts the evaporation process. A careful dynamical analysis reveals that this transition is universally determined solely by the critical exponent, and can occur within a timescale $\tau_{\rm evap}/\sqrt{S}\lesssim\tau_{\rm SC}\lesssim \tau_{\rm evap}/2$~\cite{Dvali:2020wft,Dvali:2024hsb}.\footnote{The memory burden effect is an explicit realization of the so-called \textit{quantum-breaking} of classical configurations~\cite{Dvali:2013vxa},  discussed within the framework of the BH $N$-portrait~\cite{Dvali:2012rt, Dvali:2013eja}, which questions the validity of the classical description of BHs and finds that it is violated around $\tau_{\rm sc}$.}

The probability distribution in \eqref{eq:probability} leads to a natural spread in the mass distribution of PBHs~\cite{Dvali:2024hsb}, offering a unique signature and further insights into BH evolution and mergers. While it is possible for a BH with empty memory to exist, allowing for its complete evaporation without memory burden, such cases are exponentially rare, as indicated by the probability distribution in Eq.~\eqref{eq:probability}. 

When two memory-burdened BHs merge, we expect a ``young'' BH to form, due to the large phase space volume occupied by these configurations. In fact, semiclassical BHs are realized when the memory modes are gapless and cost zero energy. At this point, the newly formed object possesses a large microstate degeneracy, and all such configurations need to be included in the amplitude, effectively saturating it~\cite{Dvali:2020wqi}. This saturation does not occur if one considers memory burdened BHs as the outcome, since BHs with different macroscopic quantum hairs have different energies and therefore do not saturate the phase space~\cite{Dvali:2024hsb}. Equivalently stated, the formation of an initially classical BH is largely insensitive to the initial pre-merger configuration, due to the large entropy involved.

A natural question is whether memory burden could dynamically induce significant deviations in the mass of the final object, compared to the initial configuration. The answer appears to be negative. In fact, as suggested by studies of mergers~\cite{Dvali:2023qlk} of memory-burdened configurations~\cite{Dvali:2024hsb}, only the merging point is altered by the memory modes, and can lead to the radiation of a small mass fraction. This, however, may result in a final BH with a spin configuration that significantly differs compared to the case without memory. Mapped to the gravitational counterparts, these results suggest that the memory burden of the two merging BHs may source significant deviations in the gravitational wave signal during the ringdown phase, with the resulting BH showing a different angular momentum than classically expected. Since Hawking radiation after the merger is sensitive to the rotation of the object, which is also determined by the initial unknown spin of the two progenitors, we restrict our analysis to non-spinning BHs. The inclusion of rotational effects would strengthen our results.

Although the merger dynamics can also lead to a partial release of the information, the newly formed BH has an area, and thus an entropy, approximately four times larger than the combined areas of the two initial BHs, providing considerable memory-storage capacity for all the initial information. Therefore, we expect that a new classical BH forms with a memory pattern governed by Eq.~\eqref{eq:probability}. 
As a consequence, post-merger the BH emits semiclassically for a time $\tau_{\rm sc}$ before stabilizing again.\footnote{The merger of extremal ultralight PBHs was discussed in Ref.~\cite{Bai:2019zcd} where it was argued that Hawking emission is reactivated. Indeed, memory-burdened PBHs  may share some phenomenological similarities with BHs that are semiclassically stabilized by extremality~\cite{Dvali:2024hsb}.} This leads to important phenomenological consequences, which are discussed next.

\subsection{High-energy particles from evaporating PBHs}

The present fraction of memory-burdened PBHs formed at temperature $T_f$ is related to their primordial abundance $\beta$ through the conservation of the PBH comoving number, as
\begin{equation}
    f_{\rm PBH} \approx 10^{21}\,\beta'\,\left(10^8{\rm\,g}/m_{\rm PBH}\right)^{1/2}\,.
\end{equation}
where $\beta' = \gamma^{1/2}\,(0.7/h)^2\,(106.75/g_{*,f})^{1/4}\beta$ is the rescaled fraction in terms of the collapsed PBH fraction $\gamma$~\cite{Carr:1975qj}, the reduced Hubble rate $h$, and the degrees of freedom at formation $g_{*,f}$. This expression generally does not apply to PBHs with masses $m_{\rm PBH} \lesssim 10^{15}$\,g, as they have already evaporated and no longer contribute to today's DM fraction $f_{\rm PBH}$. For these lighter PBHs, constraints are typically given in terms of the abundance at formation $\beta$. Here, we recast these bounds in terms of $f_{\rm PBH}$, reflecting the persistence of PBHs to present day.

The differential rate of PBH binary mergers describes a distribution of PBHs that decoupled from the Hubble flow before matter-radiation equality~\cite{Ali-Haimoud:2017rtz, Raidal:2018bbj,Liu:2018ess}. For a monochromatic mass function, the most conservative merger rate is \cite{Vaskonen:2019jpv, Hutsi:2020sol, Jedamzik:2020ypm, Young:2020scc, Jedamzik:2020omx}
\begin{equation}
    \label{eq:mergerrate}
    R_{\rm PBH}(t) \simeq\frac{5.7\times 10^{-66}}{\rm cm^3\,s}f_{\rm PBH}^{\frac{53}{37}}\left(\frac{t_0}{t}\right)^{\frac{34}{37}}\left(\frac{2 m_{\rm PBH}}{10^{10}\rm g} \right)^{-\frac{32}{37}}S\,,
\end{equation}
where the suppression factor $S$ is the product of two components: a term $S_1 \approx 0.24$ which accounts for the interactions between the binary system and the surrounding DM inhomogeneities, as well as neighboring BHs~\cite{Hutsi:2020sol}, and $S_2(x) \approx \min \left[ 1, 9.6 \times 10^{-3} x^{-0.65} \exp \left( 0.03 \ln^2 x \right) \right]$ parameterising the suppression caused by BHs absorbed by collapsed PBH clusters. Here, $x \equiv \left( t / t_0 \right)^{0.44} f_{\mathrm{PBH}}$. The function $S_2(x)$ attains a minimum at $10^{-2}$, consistently with numerical simulations~\cite{Inman:2019wvr, Franciolini:2022htd}. For light PBHs, dynamical captures induced by gravitational waves (GWs) as well as late-time dynamical capture are negligible \cite{Franciolini:2022htd}. However, local non-Gaussianity in primordial curvature perturbations can cluster PBHs at formation, potentially enhancing the merger rate by up to $\mathcal{O}(10^7)$~\cite{Raidal:2017mfl, DeLuca:2021hde, Franciolini:2022htd}. Although some level of clustering is expected from various mechanisms at the formation of the PBH binary, we consider the most conservative rate expressed by Eq.~\eqref{eq:mergerrate} as the fiducial model.

An evaporating PBH in the semiclassical phase releases a primary spectrum of light particles of species $i$ in the energy bin d$E$ at the rate d$^2N_i/{\rm d}E{\rm d}t$, depending on the multiplicity of the species, its angular momentum and spin, and the properties of the PBH. Note that the memory burden phenomenon is caused by the impossibility of the evaporation to keep track of the changing background~\cite{Dvali:2018xpy,Dvali:2020wft,Dvali:2024hsb}. Therefore, the emission spectra is determined by the mass of the BH at formation, and is constant throughout the evaporation process. A secondary emission from hadronization is also expected.

Here, we compute the differential rate for non-rotating PBHs numerically with the aid of the {\tt BlackHawk 2.3} code~\cite{Arbey:2019mbc, Arbey:2021mbl} which, for the range $m_{\rm PBH} \lesssim 10^{10}\,$g of interest here, includes all particles in the Standard Model (SM). We implemented the secondary neutrino emission following the HDMSpectra spectra~\cite{Bauer:2020jay}, which have uncertainties of at most $20\%$ for energies $E_*\sim 10^{-3}/(G\,m_{\rm PBH})$. For smaller values, the error is less controlled and may reach $\mathcal{O}(1)$.\footnote{Following the discussion of Ref.~\cite{Dondarini:2025ktz} regarding inconsistencies in the counting of primary and secondary spectra in {\tt BlackHawk}, and the handling of the secondary spectra derived by HDMSpectra, we have ensured that our results incorporate a proper treatment of these effects.}

The emission of newly formed BHs in Eq.~\eqref{eq:tausc} occurs on shorter timescales than astrophysical processes. For this, we estimate the galactic flux of particles from recently produced BHs from PBH mergers as 
\begin{equation}
    \label{eq:galflux}
    \frac{{\rm d}\Phi_{i}}{{\rm d}E}\bigg|_{\rm gal} \simeq \tau_{\rm sc}\,\int \frac{{\rm d}\Omega}{4\pi}\int{\rm d}s\,R_{\rm PBH}\,\delta(r(s,\theta))\,\frac{{\rm d^2}N_i(E)}{{\rm d}E\,{\rm d}t}\,,
\end{equation}
where $\delta(r) \equiv \rho_{\rm DM,gal}(r)/\rho_{\rm DM}$ describes the Galactic enhancement in the DM density distribution \cite{Pujolas:2021yaw}, which is here modeled as a Navarro-Frenk-White (NFW) profile~\cite{Navarro:1996gj}, $\rho_{\rm DM,gal}(r) = \rho_s (r_s/r)(1+r/r_s)^{-2}$, where we fix $r_s=15.6$\,kpc and $\rho_s$ is taken so that the DM density in the vicinity of the Solar System is $\rho_\odot=0.45{\rm\,GeV\,cm^{-3}}$, while $r_{\rm vir}$ is defined as $\delta(r_{\rm vir}) = 200$. The function $r(s,\theta)$ accounts for the line-of-sight distance $s$ of the merger from Earth, with the integration over $s$ extending up to the value for which $r=r_{\rm vir}$. Averaging the differential flux over the angular direction $\theta$ and integrating in the energy bins yields the integral flux $\Phi_i^{\rm gal}(E)$. The extra-galactic (eg) contribution to the differential flux of the species $i$ is\footnote{We thank Paolo Panci for noticing a typo in the original formula.}
\begin{equation}
    \label{eq:egflux}
     \frac{{\rm d}\Phi_{i}}{{\rm d}E}\bigg|_{\rm eg} \simeq\frac{\tau_{\rm sc}}{4\pi}\int_0^{z_f}\!{\rm d}z\left|\frac{{\rm d}t}{{\rm d}z}\right|
      R_{\rm PBH}(t(z))\frac{{\rm d^2}N_i(E')}{{\rm d}E\,{\rm d}t}e^{-\eta_i(z)}\,,
\end{equation}
where ${\rm d}t/{\rm d}z$ is the cosmological line element, $E' = E\,(1+z)$ accounts for the redshift of the energy $E'$ at emission and $z_f$ is the redshift of PBHs at formation. The opacity $\eta_i$ describes the attenuation between $z$ and today at energy $E$. More details can be found in~\cite{Dondarini:2025ktz}.

\section{Results}

\begin{figure}[ht]
    \centering
    \includegraphics[width = \linewidth]{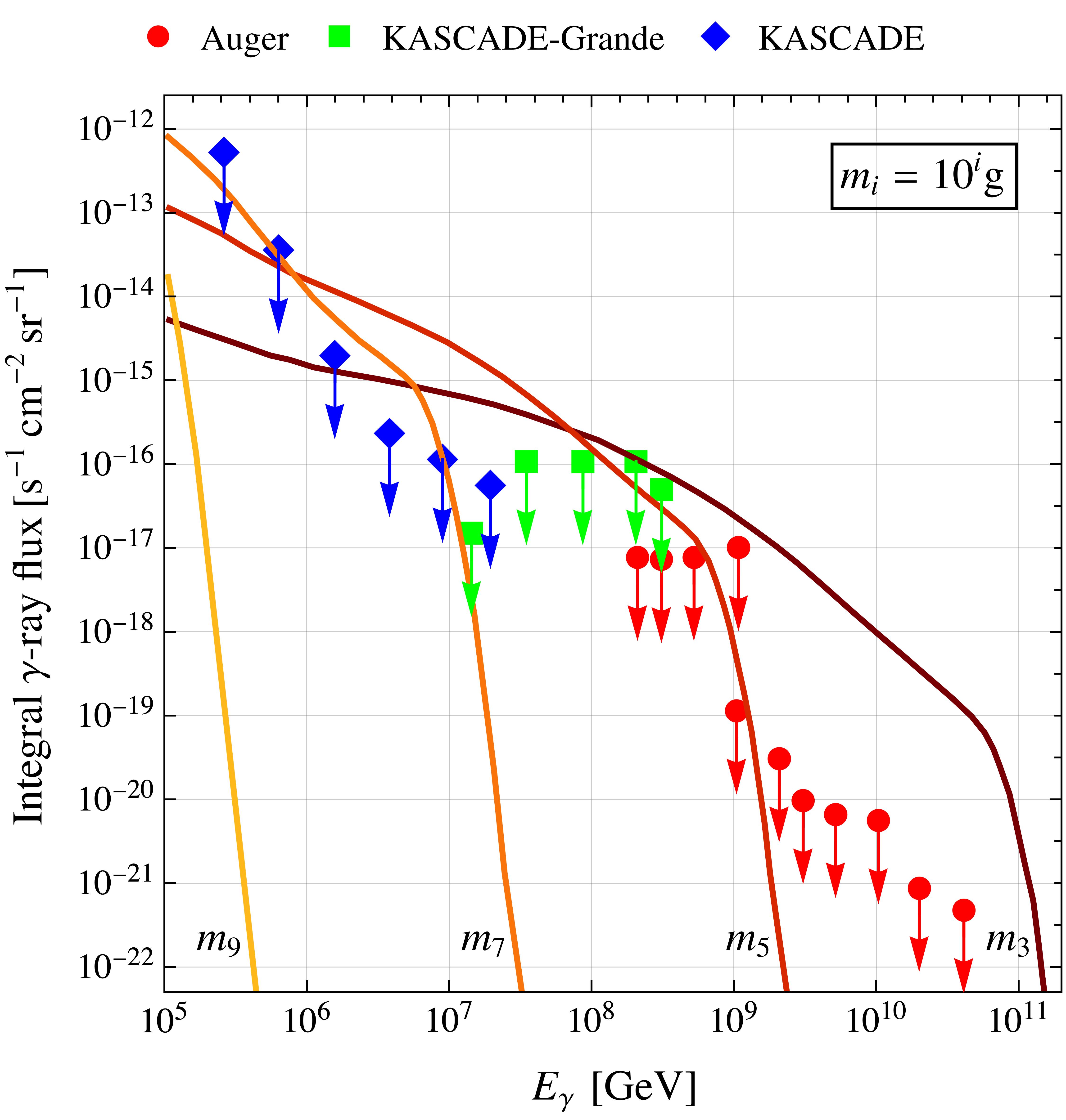}
    \caption{The integrated gamma-ray flux emitted by newly-formed BHs from memory-burdened Galactic PBH mergers, for a monochromatic PBHs distribution with $f_{\rm PBH}=1$. $m_{i}$ denotes the mass of the PBHs post-merger which is roughly twice the mass of the DM. Also shown are the upper limits at 95\% CL from Auger~\cite{Savina:2021cva, PierreAuger:2022uwd, PierreAuger:2022aty}, KASCADE, and KASCADE-Grande~\cite{KASCADEGrande:2017vwf}.}
    \label{fig:photonflux}
\end{figure}

Figure~\ref{fig:photonflux} shows the integrated gamma-ray flux $\Phi_i^{\rm gal}(E)$ in Eq.~\eqref{eq:galflux}, produced by newly-formed BHs from PBH mergers and for $f_{\rm PBH} = 1$, for different values of the PBH masses and the conservative merger rate in Eq.~\eqref{eq:mergerrate}. The absorption of gamma rays over cosmological timescales, as well as cascade effects, are implemented following Ref.~\cite{Dondarini:2025ktz}. Results are compared with the upper limits at 95\% confidence level (CL) from the Pierre Auger Observatory (Auger)~\cite{Savina:2021cva, PierreAuger:2022uwd, PierreAuger:2022aty}, the Karlsruhe Shower Core and Array Detector (KASCADE), and KASCADE-Grande~\cite{KASCADEGrande:2017vwf}. Lighter BHs are brighter due to the higher Hawking temperature and smaller entropy. For this, only the heavier BHs evade the bounds from the measured flux of gamma rays. We have not included any effects from physics beyond the SM, which would impact our results by altering the semiclassical evaporation timescale and the emission from the secondary spectrum.

\begin{figure}[ht]
    \centering
    \includegraphics[width = 1.\linewidth]{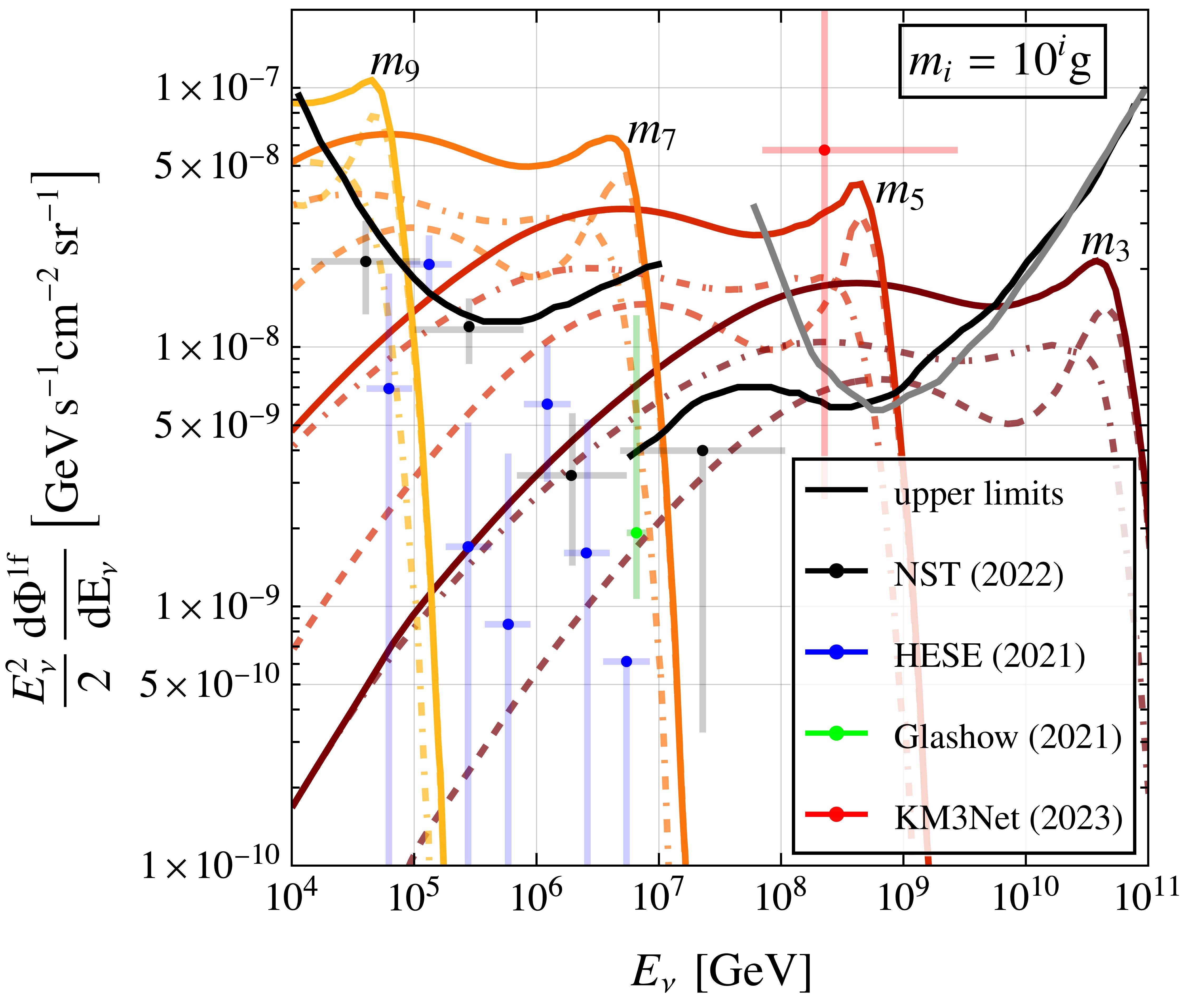}
    \caption{Per-flavor neutrino fluxes for various monochromatic PBH mass distributions with $ f_{\rm PBH} = 1 $, normalized by a factor of $2$ for better readability. Dashed (dot-dashed) lines represent the Galactic (extra-galactic) contributions, while solid lines show their sum. The black line in the upper right corner indicates the current upper limit from IceCube High-Energy Starting Events (HESE)~\cite{IceCube:2018fhm}, while the one on the left shows limits from ANTARES~\cite{ANTARES:2024ihw}. The gray line corresponds to the upper bound set by Auger~\cite{PierreAuger:2023pjg}. Black (blue) points indicate IceCube detections from Northern Sky Tracks (NST)~\cite{Abbasi:2021qfz} and HESE~\cite{IceCube:2020wum}, respectively. The green cross marks the Glashow resonance event observed by IceCube~\cite{IceCube:2021rpz}, and the red point shows the high-energy neutrino KM3-230213A detected by KM3NeT~\cite{KM3NeT:2025npi}.}
    \label{fig:neutrinoflux}
\end{figure}

We now discuss the emission of energetic neutrinos by the same sources. The differential neutrino flux per solid angle and energy is reported by the IceCube collaboration using the cascades collected over 7.5 years of operation at energies above 60\,TeV~\cite{IceCube:2020wum, IceCube:2020acn}. Below around 10\,PeV, the neutrino flux is bound by observations from ANTARES~\cite{ANTARES:2024ihw}. At energies above 10\,PeV, an upper limit at 95\% CL on the neutrino flux is reported by the IceCube High-Energy Starting Events (HESE)~\cite{IceCube:2018fhm, IceCube:2019pna} and by Auger~\cite{PierreAuger:2023pjg}.  IceCube reports candidate detections through the Northern Sky Tracks (NST)~\cite{Abbasi:2021qfz} and HESE~\cite{IceCube:2020wum}. In Fig.~\ref{fig:neutrinoflux} we show these measurements, together with the recent high-energy event detected by KM3NeT collaboration \cite{KM3NeT:2025npi} as well as the Glashow resonant observed by IceCube~\cite{IceCube:2021rpz}. We compare these experimental results with the per-flavor neutrinos differential flux  sourced by merged PBHs. The Galactic (dashed), extragalactic (dot-dashed), and total (solid) contributions are obtained assuming the merger rate in Eq.~\eqref{eq:mergerrate}.

\section{Discussion}

Assuming that a fraction $f_{\rm PBH}$ of the DM relic abundance consists of light PBHs, we derive a constraint on the model in the $(m_{\rm PBH}, f_{\rm PBH})$ plane by performing a least-square analysis that accounts for the neutrino flux measurements and upper bounds reported by the IceCube collaboration~\cite{IceCube:2020wum, IceCube:2020acn, IceCube:2018fhm, IceCube:2019pna}. 

\begin{figure}[ht]
    \centering
    \includegraphics[width = 1.\linewidth]{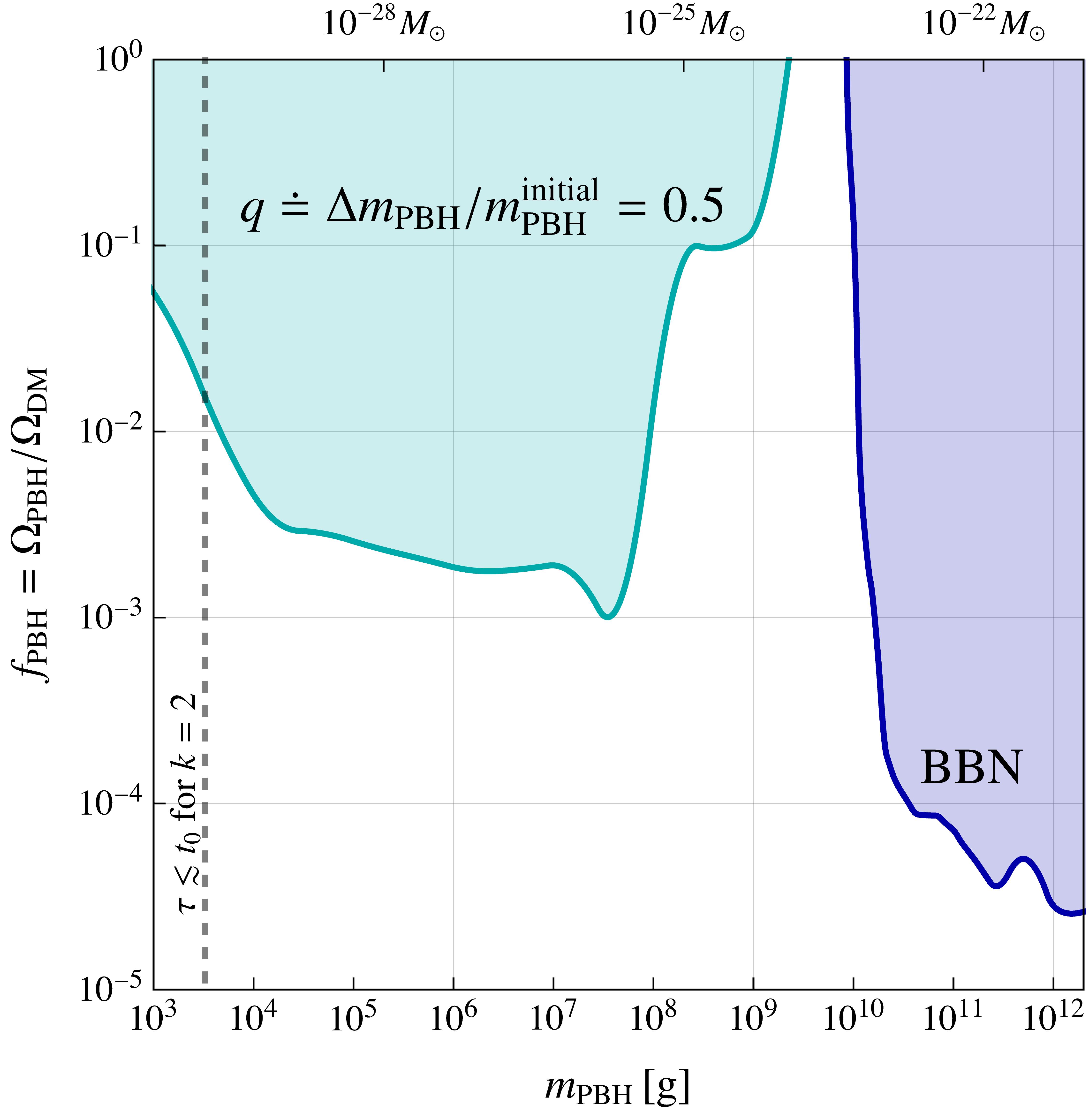}
    \caption{The DM fraction in PBHs of mass $m_{\rm PBH}$ as a function of the PBH mass. The results in this work consist in the bound from IceCube (filled cyan area). Also shown are bounds from BBN considerations, taken from~\cite{Carr:2009jm,Carr:2020gox,Thoss:2024dkg}. In deriving the bound, we assume that the memory burden effect occurs when the PBH has lost half of its initial mass. Different mass-loss fraction are possible, potentially altering our findings as discussed in the text. 
    }
    \label{fig:BHbound}
\end{figure}

Using the merger rate in Eq.~\eqref{eq:mergerrate}, Fig.~\ref{fig:BHbound} excludes light PBHs of masses $ m_{\rm PBH}\lesssim  10^9\,\rm g$ as the totality of the DM (filled cyan area). 
The mass range is further constrained, with the upper bound $m_{\rm PBH}\lesssim 10^{10}$\,g from considerations over the effects of PBH evaporation at the onset of Big-Bang nucleosynthesis (BBN, blue contour)~\cite{Carr:2009jm,Carr:2020gox,Thoss:2024dkg}. Also shown is the value $m_{\rm PBH} \sim 3\times 10^3$\,g  (vertical dashed line), for which $\tau_{\rm sc
}\simeq t_0$ when $k=2$, c.f., Eq.~\eqref{eq:lifetimeenhancement}. For $k= 2$, constraints have been placed on PBH masses up to approximately $m_{\rm PBH}\sim 10^5$\,g from the emission of photons~\cite{Alexandre:2024nuo, Thoss:2024hsr} or neutrinos~\cite{Chianese:2024rsn} (see \cite{Dondarini:2025ktz} for a complete analysis). However, the signal following the merger considered here is not sensitive to $k$, offering a novel and complementary signature that does not depend on the specific details of the stabilized phase.  

Our results suggest that the window where memory-burdened PBHs constitute the DM is in tension with current measurements on the neutrino flux.  
As shown in Fig.~\ref{fig:BHbound}, only PBHs in the mass range $10^9 \lesssim m_{\rm PBH}/{\rm g} \lesssim 10^{10}$ could make up a sizable fraction of the DM. Obviously, it is possible  that memory burden sets in before half of the mass has been emitted semi-classically. This possibility is actually supported in theoretical studies, where it has been found that memory burden may emerge after an $\mathcal{O}(1/\sqrt{S})$ fraction of the initial mass has been emitted~\cite{Dvali:2018xpy, Dvali:2018ytn, Dvali:2020wft, Alexandre:2024nuo, Dvali:2024hsb}. Independently motivated studies found that, on comparable timescales, the semiclassical description of BHs ceases to be accurate~\cite{Dvali:2011aa,Dvali:2012rt, Dvali:2012en, Dvali:2012wq, Dvali:2013vxa, Dvali:2015wca,Michel:2023ydf}. In such scenario, the bounds presented here would no longer apply.

Comparable, and potentially stronger constraints are expected in a similar mass range due to gamma-ray flux measurements by the Large High Altitude Air Shower Observatory (LHAASO)~\cite{LHAASO:2019qtb}, which detected gamma-rays above about 10\,TeV. Studies of gamma and cosmic ray emissions from PBH clusters could further constraints $f_{\rm PBH}$, particularly for masses smaller than $10^5$\,g, while cosmic ray anisotropies might reveal nearby PBH subhalos~\cite{IceCube:2010tuw, IceCube:2011fxx, HAWC:2014ics, HAWC:2018wju, TibetASg:2021kgt}, though more data is needed~\cite{Giacinti:2011mz}. For higher-mass PBHs, upcoming neutrino observatories like IceCube-Gen2 radio array~\cite{IceCube-Gen2:2021rkf}, the Radio Neutrino Observatory in Greenland (RNO-G)~\cite{RNO-G:2020rmc}, and the Giant Radio Array for Neutrino Detection (GRAND)~\cite{GRAND:2018iaj}, will probe neutrinos at energies above 10\,PeV with improved sensitivity. In this direction, we ought to mention the recent measurement of the high-energy - above $10^2\,$PeV - neutrino event detected by the KM3NeT collaboration~\cite{KM3NeT:2025npi}. These observatories could constrain $f_{\rm PBH}$ by analyzing neutrino fluxes, especially when combined with data from Galactic sources, improving detection reach by $\mathcal{O}(10)$ over 10 years. Directional neutrino measurements will be crucial for distinguishing PBH contributions, providing a test for this scenario. Evaporating PBHs may emit gravitons, peaking at frequencies $\gtrsim 10^{30}$\,Hz for $m_{\rm PBH} \lesssim 10^8$\,g, beyond the reach of current GW detectors~\cite{Domcke:2022rgu, Schmieden:2023fzn, Gatti:2024mde}. GW abundance estimates suggest $\Omega_{\rm GW} \lesssim 10^{-12}$ at the peak. 

\section{Conclusions}

In this paper, we have explored the astrophysical implications to a PBH population affected by the \textit{memory burden} effect. Due to the associated halt in the evaporation process, light PBHs might live over cosmological timescales and constitute viable DM candidates, undergoing binary merging and forming new BHs of similar mass at much later cosmic times than their formation epochs. We argued that newly-formed BHs initially do not carry memory burden and can therefore emit energetic particles, in the form of cosmic rays, gamma rays, and neutrinos, which can be searched at dedicated facilities. We have estimated the integral spectrum in gamma rays from Galactic mergers and the differential spectrum of per-flavor neutrinos. Based on the current available measurements by IceCube, we have derived bounds on the PBH fraction. We found that, for a late memory burden onset, after the BH has lost roughly half of its initial mass, PBHs can account for the entirety of DM only in a small window $10^9 \lesssim m_{\rm PBH}/{\rm g} \lesssim 10^{10}$. Our findings are weakened if the memory burden sets in at parametrically earlier times. However, if the merger rate is increased by primordial non-Gaussianity, the mass range that is potentially excluded sensibly widens and our constraints strengthen. Future detectors will be able to probe even deeper into the remaining viable parameter space. Additional future directions include the careful study of the model with cosmic-ray and gamma-ray signatures from Galactic substructures.

{\bf Note Added.} After the completion of this work, Ref.~\cite{Dondarini:2025ktz} performed an in depth analysis of neutrinos, gamma-rays and early Universe signatures (CMB) resulting from the presented mechanism. We refer the interested reader to~~\cite{Dondarini:2025ktz} for a detailed discussion of the existing constraints as well as the characterization of the duration of the semiclassical emission phase (given by $q$ in \eqref{eq:tausc}) that allows $f_{\rm PBH}=1$ in the entire mass window. Our result in Fig.~\ref{fig:BHbound} is qualitatively consistent to theirs. Some minor discrepancies follow from the fact that~\cite{Dondarini:2025ktz} adopted slightly different datasets and included the energy binning of the experimental points in the analysis.

\vspace{.1cm}
\begin{acknowledgments}
We thank Gia Dvali for useful discussions and for reading the manuscript. We further thank Gwenael Giacinti for useful comments. Finally, we are grateful to Concha Gonzalez-Garcia for correcting a previous mistake in the detectability of the neutrino mass nature at IceCube and for further discussions. MZ is thankful to Alessandro Dondarini, Giulio Marino and Paolo Panci for useful ongoing discussion on related topics. We acknowledge support by the National Natural Science Foundation of China (NSFC) through the grant No.\ 12350610240 ``Astrophysical Axion Laboratories''. LV acknowledges support by the Istituto Nazionale di Fisica Nucleare (INFN) through the ``QGSKY'' Iniziativa Specifica project. LV thanks the Tsung-Dao Lee Institute for its hospitality during the final stages of this work.
\end{acknowledgments}

\setlength{\bibsep}{4pt}
\bibliography{citations.bib}

\end{document}